\documentclass[12pt,letterpaper,aps,nofootinbib,floats,floatfix]{revtex4}
\usepackage{graphicx}
\usepackage{hyperref}
\usepackage{verbatim}
\usepackage{color}

\begin{document}

\newcommand {\be} {\begin{equation}}
\newcommand {\ee} {\end{equation}}
\newcommand {\bea} {\begin{eqnarray}}
\newcommand {\eea} {\end{eqnarray}}
\newcommand {\eq} [1] {eq.\ (\ref{#1})}
\newcommand {\Eq} [1] {Eq.\ (\ref{#1})}
\newcommand {\eqs} [2] {eqs.\ (\ref{#1}) and (\ref{#2})}
\newcommand {\fig} [1] {fig.\ \ref{#1}}
\newcommand {\Fig} [1] {Fig.\ \ref{#1}}
\newcommand {\figs} [2] {figs.\ \ref{#1} and \ref{#2}}

\def \eps {\epsilon}
\def \veps {\varepsilon}
\def \pl {\partial}
\def \hf {{1 \over 2}}
\def \mf {\mathbf}
\def\figdir{}
\def\sss{\scriptscriptstyle}

\def\lsim{\mbox{\raisebox{-.6ex}{~$\stackrel{<}{\sim}$~}}}
\def\gsim{\mbox{\raisebox{-.6ex}{~$\stackrel{>}{\sim}$~}}}

\title{\bf Dynamical Fine Tuning in Brane Inflation}

\author{
\normalsize James~M.~Cline, Loison~Hoi and Bret~Underwood }
\affiliation{
Department of Physics, McGill University\\
3600 University Street, Montr\'eal, Qu\'ebec, Canada H3A 2T8\\
E-mail: jcline@physics.mcgill.ca, hoiloison@physics.mcgill.ca, bjwood@physics.mcgill.ca}
\date{15 May 2009}

\begin{abstract}

We investigate a novel mechanism of dynamical tuning of a flat potential
in the open string landscape within the context of
warped brane-antibrane inflation 
in type IIB string theory.  
Because of competing effects between interactions with the moduli
stabilizing $D7$-branes in the warped throat and anti-$D3$-branes at the tip,
a stack of branes gives rise to a local minimum
of the potential, holding the branes high up in the throat. 
As branes successively tunnel out of the local minimum to the bottom
of the throat the potential barrier becomes lower
and is eventually replaced by a flat inflection point, around
which the remaining branes easily inflate.  This dynamical flattening
of the inflaton potential reduces the need to fine tune the potential
by hand, and also leads to successful inflation for a larger range
of inflaton initial conditions, due to trapping in the local minimum.

\end{abstract} 

\maketitle

\section{Introduction}
\label{sec:introduction}

The suitability of a scalar field potential for inflation is sensitive
to the form of nonrenormalizable
interactions: even if their effects are Planck-suppressed, a
correction to the inflaton mass of order $V/M_p^2$ is enough to
spoil inflation.  Therefore 
 it makes sense to use ultraviolet (UV) complete theories such as
string theory to compute the form of, and possible corrections to,
inflationary potentials and models.
Important progress has been made in the last few years in refining string
theoretic models of inflation, especially those based on the motion
of a mobile $D3$ brane in the extra, compact dimensions
\cite{DvaliTye,KKLMMT,Baumann,Uplifting,DelicateUniverse,Chasing,Panda1,Panda2,HolographicSystematics}.  
Increasingly, assumptions
about the form of the inflaton potential that were necessary to 
make initial progress are being replaced by
calculations of the potential from effective field theory 
and AdS/CFT techniques arising from string theory constructions.

\begin{figure}[htp]
\centerline{
\includegraphics[scale=.25]{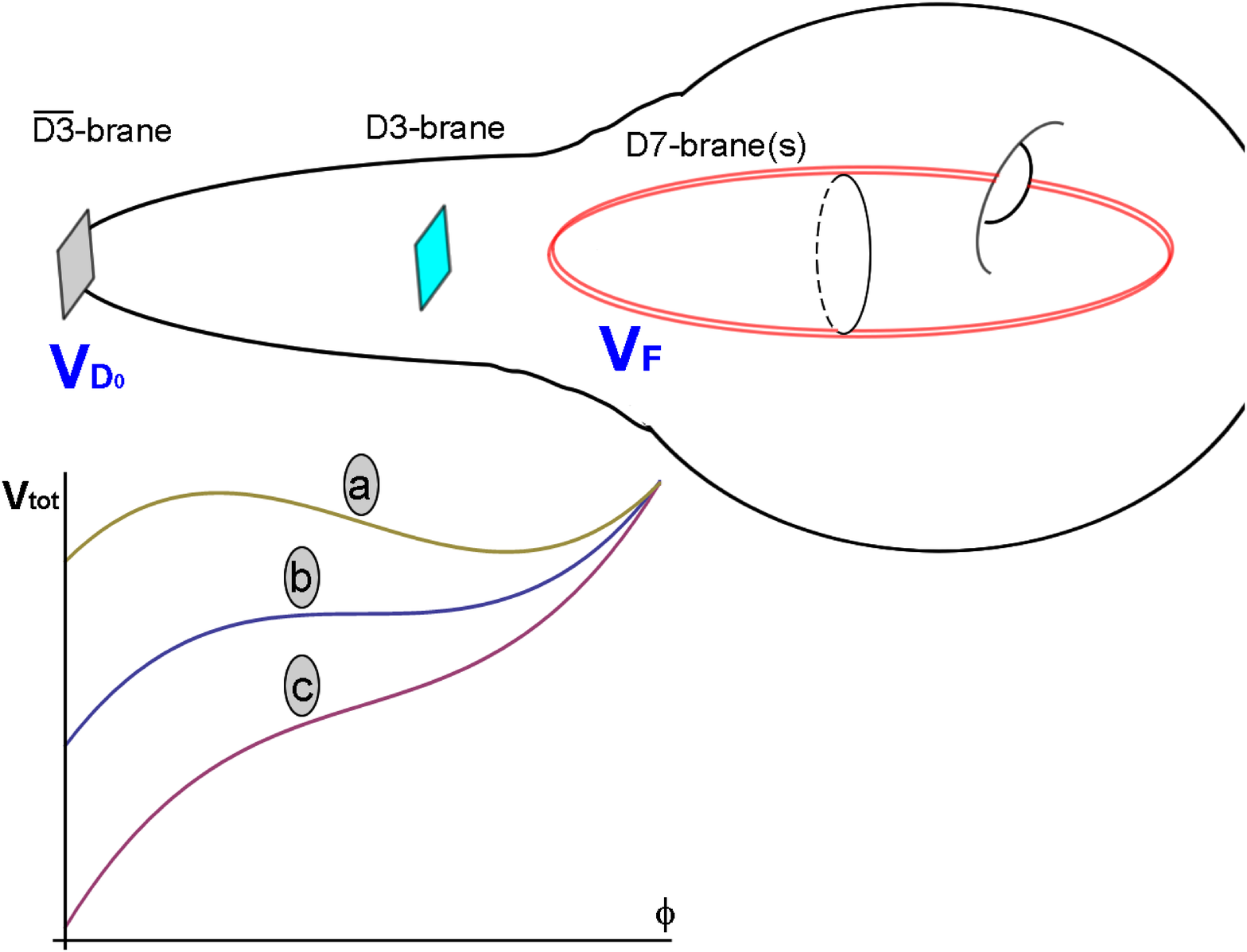}}
\caption{\small Warped brane inflation suffers from a fine tuning problem 
due to the fact that small changes in parameters
(a)-(c) lead to drastic changes in the behavior of the potential which 
make it unsuitable for inflation.}
\label{fig:BranesInThroat_FineTuning}
\end{figure}

Despite this progress (or rather because of it), several 
fine-tuning challenges 
typical in usual inflationary model building which can be sensitive to UV-physics
still remain, namely
fine-tuning of the precise functional form of the potential and of
the initial conditions of the inflaton field.
In Figure \ref{fig:BranesInThroat_FineTuning} we illustrate the problem of functional
fine-tuning for the brane inflationary scenario of \cite{DelicateUniverse}: for
one set of parameter values, the potential has an inflection point
which is flat enough to support a sufficient number of e-folds of
inflation (labeled (b)), while small deviations from these parameters
((a) and (c)) can lead to a potential which 
does not support inflation.

In addition to the functional fine tuning of the potential, small
field inflation models such as brane inflation in which the inflaton 
ranges over sub-Planckian field values typically also suffer from
an {\it overshoot problem}.  This occurs when the initial conditions
are such that the inflaton has too much kinetic energy when it
enters the inflationary region to satisfy the slow roll
conditions, and it overshoots the slow roll region (for a recent discussion
of the overshoot problem in string theory models of inflation, see
\cite{AccidentalInflation,Itzhaki:2007nk,AttractiveBrane,Itzhaki:2008ih}).
This problem is particularly acute for the inflection-point type
of inflationary potentials which arise from the brane constructions 
of \cite{DelicateUniverse}.

{}From the point of view of the 4-dimensional effective theory, 
the fine tuning in both the potential and initial conditions must
be done by hand. However, since the potential originates from a higher 
dimensional string theoretic model, it is interesting to investigate whether
this tuning can arise in a dynamical way.  One method of dynamical
tuning commonly used in string theoretic constructions is that
of the landscape; as one moves, by tunneling or rolling, in the 
multidimensional field space the parameters of the theory
(such as the fluxes wrapped on cycles in the internal space) 
change in a dynamic way. 

In this paper we will consider a mechanism of dynamical resolution 
of these fine tuning problems in the open string landscape; we will 
focus on the specific 
model in which brane inflation arises as the interaction
between $D3$- and $\overline{D}3$-branes in a locally warped throat
in the presence of moduli stabilizing ingredients such as $D7$-branes
\cite{DelicateUniverse}, as shown schematically in 
Figure \ref{fig:BranesInThroat_FineTuning}.
Recently it was shown that the (relative) functional fine tuning 
problem in this model is
greatly ameliorated for certain optimal values of the parameters
\cite{DBraneFineTuning}: near this optimal region, each parameter
could vary at the level of a part in a few without spoiling
inflation. 
Nevertheless, once the full allowed field range is included
correctly the model still requires fine tuning of the initial
conditions to avoid overshooting the inflationary region
of the inflection point.

\begin{figure}[htp]
\centerline{
\includegraphics[scale=.25]{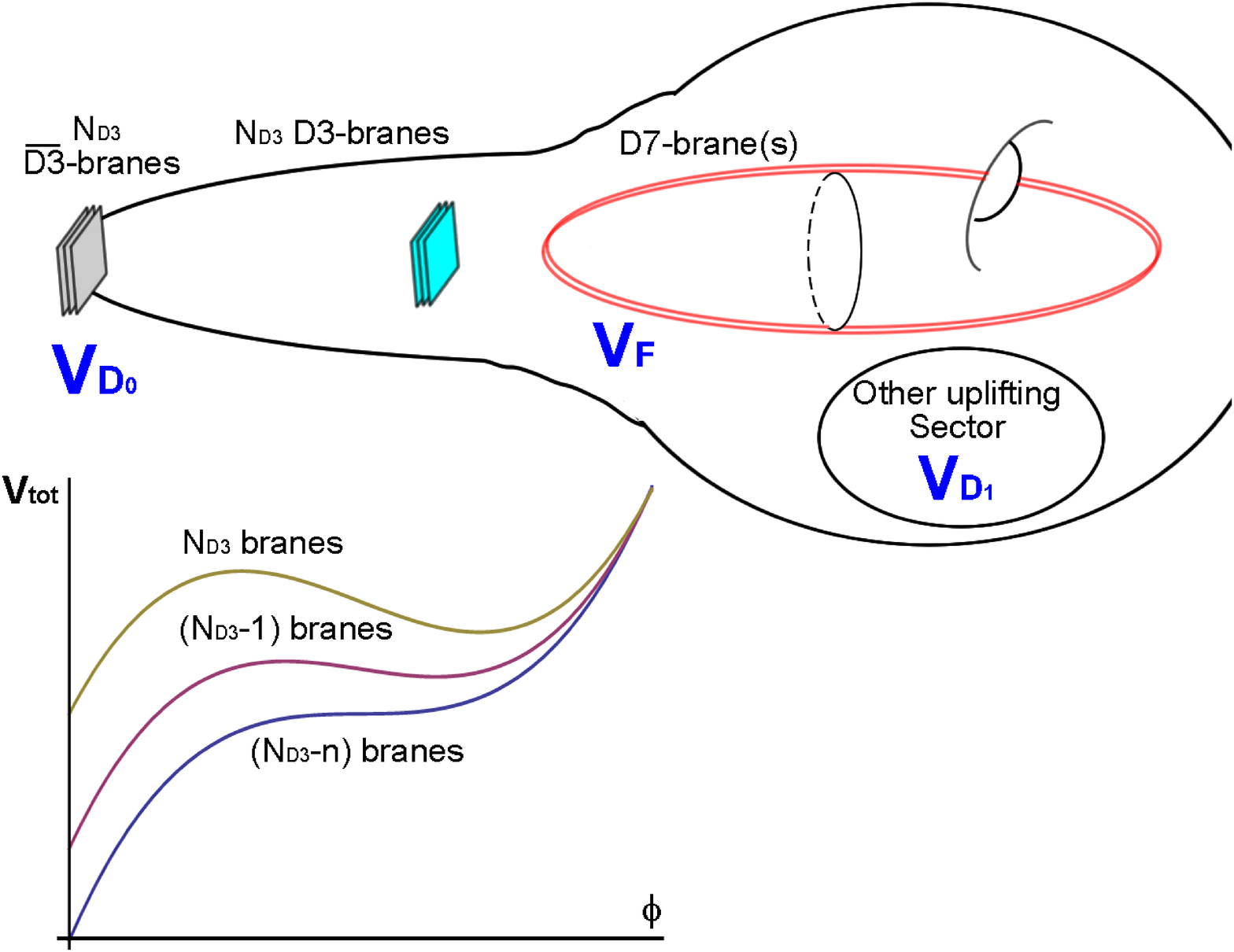}}
\caption{\small In the presence of multiple branes and antibranes, the inflationary potential
develops a local minimum which traps the branes at a finite distance in the throat.  As individual
branes tunnel out of the local minimum the potential barrier decreases, until the potential
forms a flat inflection point.  This dynamical process can aid in reducing the amount of necessary fine-tuning
of the parameters in the potential, and can also help to reduce the overshoot problem.}
\label{fig:BranesInThroat}
\end{figure}

Revisiting a novel mechanism, first suggested in 
\cite{MultibraneFlattening}, 
we will show
that in the presence of sufficiently many branes and anti-branes
the potential typically develops a local minimum that serves to trap
the branes at a finite distance in the throat.  The local minimum
arises because of a competition between the backreaction of the
$D3$-branes on the four cycle volume and the attractive force of the
antibranes at the tip of the throat. Individual branes have a finite
probability to tunnel out of the local minimum and annihilate with
the antibranes at the tip, and as they do so they change the balance
of forces on the remaining branes. As the number of trapped branes
decreases so does the height of the potential barrier of the local
minimum, until at some point the local minimum disappears and is
replaced with an inflection point potential.  This scenario is shown
schematically in Figure \ref{fig:BranesInThroat}.

The tunneling of the branes and subsequent modification to the
potential is equivalent to scanning over parameters of the potential
in a dynamical way.  As long as other sectors dominate the uplifting
energy, the step size can be sufficiently small and we can
dynamically access the parameters needed to obtain a sufficient
amount of inflation, alleviating the parameter fine tuning problem. 
Moreover, prior to the inflationary period, as mobile $D$-branes fall
into the throat they are naturally trapped in the local minimum, both
due to requirement of large enough kinetic energy to overcome the potential barrier
as well as due to energy loss in collisions with other trapped
$D$-branes \cite{BeautyAttractive,McAllister:2004gd}.  This leads to a 
relaxation of
the overshoot problem.
The original scenario \cite{MultibraneFlattening} was
incomplete because 
important stringy
corrections to the nonperturbative superpotential for the $D3$ brane
position in the throat were not yet known.  
Our new observation is that the dynamical tuning 
mechanism does in fact exist when these corrections are taken
into account.

The paper is organized as follows.  In Section \ref{sec:Potential} we review
the form of the inflationary potential, and illustrate the functional 
and initial conditions fine-tuning problems.  In Section \ref{sec:Dynamical}
we demonstrate the mechanism in which the potential is dynamically flattened,
and discuss how this can help in relaxing the amount of functional and initial
conditions fine-tuning for this model.  We also discuss the constraints 
on the parameters of the model for the rate of the dynamical tunneling process
to be faster than other decay rates, and find these constraints to be quite
strong.
Finally, in Section \ref{sec:Conclusion} we summarize our findings.

\section{Potential for D-brane Inflation}
\label{sec:Potential}

The model of type IIB $D3$-brane inflation we will consider consists of several 
ingredients \cite{DvaliTye,KKLMMT,DelicateUniverse,Chasing,Uplifting}: 
\begin{itemize}
\item Stacks of spacetime filling $D3$- and $\overline{D3}$-branes, 
where the radial separation between the branes
serves as the inflaton field and the energy from the non-BPS $\overline{D3}$-branes provide
the inflationary energy.  
In particular, for simplicity we will consider $N_{D3}$ mobile $D3$-branes and the
same number of $\overline{D3}$-branes.  If this is not the case then the energy from 
reheating is trapped on the remaining branes at the tip, so reheating of the standard
model sector does not occur unless the standard model itself is realized in this throat,
and it is not known whether this is possible.  In contrast, if there are no branes remaining
in the inflationary throat after inflation then the inflationary energy is channeled
into closed string modes \cite{Barnaby:2004gg,WarpedReheating,hep-th/0507257,hep-th/0602136,arXiv:0710.1299}
that may reheat the standard model, which can be constructed
elsewhere in the compact space.  
Our results, however, 
will not be sensitive to this assumption because it only changes the dependence
on the number of $D3$-branes in the subdominant Coulomb attraction,
and it is straightforward to generalize the scenario.
\item A warped throat, which suppresses the attractive Coulombic force between
the $D3-\overline{D3}$ pair and provides a local metric of the internal space in which to do concrete computations.
For concreteness, we will be considering the warped deformed conifold \cite{KS} glued to a compact
space as described in \cite{GKP}.
\item $D7$-brane(s) wrapped on a four cycle $\Sigma_4$ extended along the radial direction of
the throat (and extending into the bulk),
upon which there are non-perturbative effects such as gaugino condensation (alternatively, the $D7$-brane
can be replaced by a Euclidean $D3$-brane instanton) necessary for the 
stabilization of the K\"ahler modulus of the internal geometry. 
\item Other
sectors elsewhere in the bulk geometry such as $\overline{D3}$-branes in other warped throats or
fluxes on $D7$-branes \cite{DTermsBurgess,DTermsJockers,DTermsCremades} which contribute to the uplifting energy.
\end{itemize}
The schematic picture of this setup is shown in Figure \ref{fig:BranesInThroat}.
We will assume that contributions to the inflaton potential from the bulk (as recently
computed in \cite{HolographicSystematics} and discussed further in \cite{ChenScanning}) 
are subdominant in comparison to the
force on the $D3$-branes from the part of the $D7$-brane in the throat.
Because each of the $N_{D3}$ branes in the stack are all subject to the same forces, we will
identify the inflaton field with the the overall position modulus $z^\alpha$ 
of the stack as a whole.

The potential can be computed using the formalism of ${\mathcal N}=1$ supergravity,
\begin{equation}
V = V_F + V_D = e^{{\mathcal K}} \left[K^{A\bar{B}}D_A W D_{\bar{B}}\bar{W}-3|W|^2\right] + V_D
\label{eq:VSugra}
\end{equation}
where the superpotential and K\"ahler potential\footnote{Recent progress in computing
effective theories in strongly warped backgrounds can be found in \cite{dg,gm,stud,dt}; in particular,
the K\"ahler potential for the universal K\"ahler modulus, including generic warp factor corrections,
was recently computed in \cite{ftud}.  We will shift the warping corrections of \cite{ftud} into
the (undetermined) non-perturbative superpotential coefficient $A_0$.} are given by
\begin{eqnarray}
W &=& W_0 + A(z^\alpha) e^{-a \rho} \\
{\mathcal K} &=& - 3\log[\rho+\bar{\rho}-\gamma k(z^\alpha,\bar{z}^{\bar{\alpha}})] \equiv - 3 \log U(\phi,\sigma)
\end{eqnarray}
in which $k(z^\alpha,\bar{z}^{\bar{\alpha}})$ is the little K\"ahler potential of the internal
metric $g_{\alpha\bar{\beta}} = \partial_\alpha\partial_{\bar{\beta}} k$,
$W_0$ represents the contribution from the GVW flux-induced superpotential stabilizing the
complex moduli \cite{GVW,GKP} 
\begin{equation}
W_0 \equiv \left(\int G_3\wedge\Omega\right)\, ,
\end{equation}
and we took a single K\"ahler modulus $\rho = \sigma + i \varsigma $ (the details of the 
stabilization of $\varsigma$ will not be important as its vev can be compensated by the phase
of $A(z^\alpha)$).
The presence of $N_{D3}$ $D3$-branes leads to backreaction on the volume $\Sigma_4$ of the moduli-stabilizing
$D7$-branes \cite{Baumann} which induces a dependence of the non-perturbative superpotential on the
position of the $D3$-brane stack
\begin{equation}
A(z^{\alpha}) \equiv A_0 g^{N_{D3}/n}(z^\alpha)
\end{equation}
where $A_0$ depends on the details of the gaugino condensation and stabilized values
of the complex structure moduli, $n$ is the number of $D7$-branes, and $g(y^\alpha) = 0$
is the embedding equation of the four cycle $\Sigma_4$.  The dependence
on the number of $D3$ branes in the stack can be inferred by noticing 
that $A(z^\alpha) \sim e^{\delta V_w(z^\alpha)}$
where $\delta V_w \sim N_{D3} \ln g(z^\alpha)$ is the perturbation of the warped four cycle
volume \cite{Baumann} from the presence of the $D3$-branes, which is linear in $N_{D3}$.
For concreteness, we will choose the Kuperstein embedding \cite{Kuperstein}, for which the
embedding as a function the radial coordinate $\phi$ is $g(\phi) = 1 + \left(\frac{\phi}{\phi_\mu}\right)^{3/2}$,
where $\phi_{\mu}$ is a parameter in the embedding of the D7-brane which determines how far it reaches
into the throat.

We will express the potential in terms of the following useful dimensionless rescalings of the 
radial\footnote{We have stabilized the additional angular
directions of the mobile D-brane(s) throughout the inflationary period as in \cite{DelicateUniverse,Chasing,Uplifting},
although their presence may lead to additional effects at the end of inflation 
\cite{MultiDBIHuang,MultiDBILanglois1,MultiDBILanglois2,MultiDBIChen}.} position
of the $D3$-brane stack $\phi$ and the K\"ahler modulus $\sigma$,
\begin{eqnarray}
\label{phimu}
x & \equiv & \sqrt{N_{D3}}\frac{\phi}{\phi_{\mu}}\\
\omega &\equiv & a\sigma = \frac{2\pi}{n} \sigma\, .
\end{eqnarray}
The factor of $\sqrt{N_{D3}}$ in (\ref{phimu}) simply comes from the 
rescaling of $N_{D3}$ identical kinetic terms.  With this rescaling,
the stack of $D3$-branes reaches the $D7$-branes at $x_{\rm max} \equiv \sqrt{N_{D3}} > 1$.
The stabilized value of the K\"ahler modulus 
\begin{equation}
\label{eq:w*}
\left.\frac{\partial V}{\partial \omega}\right|_{\omega=\omega_*(x)} = 0
\end{equation}
depends on the $D3$-brane position $\omega_* = \omega_*(x)$, and as such
cannot be integrated out of the
effective four dimensional potential \cite{DelicateUniverse,Chasing,Uplifting,Panda1,DBraneFineTuning}.

It is useful to define several additional parameters to compute the inflationary potential.
We will define the stable value of the K\"ahler modulus before uplifting $\omega_F$ by solving
\begin{equation}
3 \left|\frac{W_0}{A_0}\right|e^{\omega_F} = 2\omega_F + 3\, ,
\end{equation}
thus the parameter $W_0$ can be traded for $\omega_F$.  Similarly, we define the stable value
of the K\"ahler modulus after inflation $\omega_0$ as,
\begin{equation}
\left.\frac{\partial V}{\partial \omega}\right|_{x=0,\omega=\omega_0} = 0\, .
\end{equation}
It is easy to show that requiring the vacuum energy at the end of inflation to be equal
to the current cosmological constant (which is approximately zero for our purposes) $V(0,\omega_0) \approx 0$
leads to the relation
\begin{equation}
(2\omega_F+3)e^{\omega_0-\omega_F}-2\omega_0 - 5 = 0\, ,
\end{equation}
thus $\omega_0 = \omega_0(\omega_F)$ is not a free parameter.
Another useful parameter is the ratio of the D-term and F-term energies,
\begin{equation}
s \equiv \frac{V_D(0,\omega_F)}{|V_F(0,\omega_F)|}\, .
\label{eq:sDef}
\end{equation}
The constraint on uplifting at the end of inflation leads to a relation \cite{DBraneFineTuning},
\begin{equation}
s = \frac{\omega_0(\omega_F)+2}{\omega_F}\left(\frac{2\omega_F+3}{2\omega_0(\omega_F)+5}\right)^2
\cong\frac{\omega_F}{\omega_0(\omega_F)}\left[1+3\left({1\over\omega_F}-{1\over\omega_0(\omega_F)}\right)\right]\, ,
\end{equation}
thus $s$ is also not a free parameter and is fixed to $s\approx 1$ ($\omega_0\cong\omega_F\gg1$) by the 
requirement $V(0,\omega_0) \approx 0$.

Writing the potential in terms of these parameters, we 
have\footnote{The extra dependence upon $N_{D3}$ can be inferred from
ref.\ \cite{DelicateUniverse} as follows. Before doing any rescaling
of $\phi$ due to the factor of $N_{D3}$ in the kinetic term, 
the parameter $\gamma$ in eq.\ (2.10) of that paper is rescaled by 
$N_{D3}$.  The quantity $\alpha_z$ of eq.\ (3.4) also gets rescaled by
$N_{D3}$, coming from differentiating $g^{N_{D3}/n}$.  Taken together,
these imply that the two terms in the F-term potential which depend
upon derivatives with respect to the brane modulus (the last line of 
our eq.\ (\ref{eq:PotentialFTerm})) both get multiplied
by $N_{D3}$.  Finally we rescale $x\to x/\sqrt{N_{D3}}$ to canonically
normalize the inflaton kinetic term.}
\begin{eqnarray}
\label{eq:PotentialFTerm}
V_F(x,\omega) &=& \frac{a|A_0|^2}{3 U^2(x,\omega)} e^{-2\omega} 
g(\frac{x}{\sqrt{N_{D3}}})^{2N_{D3}/n}\nonumber\\
&&\times\left[2\omega+6-2 (2\omega_F+3)e^{\omega-\omega_F}g(\frac{x}{\sqrt{N_{D3}}})^{-N_{D3}/n}\right.\nonumber\\
&&\left.+ \frac{3N_{D3}}{n g(\frac{x}{\sqrt{N_{D3}}})} \left(\frac{c x}{N_{D3}^{1/2}g(\frac{x}{\sqrt{N_{D3}}})}-\left(\frac{x}{N_{D3}^{1/2}}\right)^{3/2}\right)\right] 
\\
V_D(x,\omega) &=& \frac{1}{U^2(x,\omega)} \left(D_1 + \frac{2 N_{D3}
T_0}{\begin{displaystyle}1+C_D \frac{2N_{D3}^3 T_0}{x^4}\end{displaystyle}}\right) \nonumber \\
	&=& \frac{2a|A_0|^2}{3 U(x,\omega)^2} s\omega_F
e^{-2\omega_F}\left(1+\frac{D_{01}(N_{D3})}{\begin{displaystyle}1+C_D \frac{2N_{D3}^3 T_0}{x^4}\end{displaystyle}}\right)
\label{eq:PotentialDTerm}
\end{eqnarray}
where $c = 9/(4n\omega_0\phi_\mu^2)$, $C_D = (27/64\pi^2)\phi_\mu^{-4}$, and we defined,
\begin{equation}
D_{01}(N_{D3}) = 2\frac{N_{D3} T_0}{D_1}\, ,
\label{eq:D01def}
\end{equation}
as the ratio of the uplifting energy from the warped $D3$-$\overline{D3}$ interaction and the uplifting energy 
left over after inflation.
The definition of $s$ in (\ref{eq:sDef}) allows us to determine the leftover uplifting energy
in terms of the F-term potential,
\begin{equation}
D_1 = \frac{2}{3} a |A_0|^2 s\omega_F e^{-2\omega_F}\, .
\label{eq:UpliftCondition2}
\end{equation}
The K\"ahler potential function in these variables is
\begin{equation}
U(x,\omega) = \frac{1}{a} \left(2\omega - \frac{1}{3}\omega_0 
\phi_\mu^2 x^2\right)\, .
\end{equation}
Notice that the $N_{D3}$ dependence from the
rescaling (\ref{phimu}) cancels out in the K\"ahler potential.

Altogether, we find that the potential depends on the set of (effectively) 
continuous\footnote{In principle all parameters should be determined by a set of
discrete choices of fluxes and branes; in practice it is not always clear how to
express certain quantities in terms of fluxes (e.g.\ $A_0$), and so we will
treat the discrete steps of these quantities as small enough that they are
effectively continuous for our usage.}
and discrete parameters
\begin{eqnarray}
\label{eq:Parameters}
\{D_{01},\omega_F,A_0,\phi_\mu\} & & \mbox{effectively continuous} \\
\{n,N_{D3}\} & &\mbox{discrete}\, .\nonumber
\end{eqnarray}

\subsection{Functional Fine-Tuning Problem}

As the parameters in (\ref{eq:Parameters}) are varied the potential (\ref{eq:PotentialFTerm}-\ref{eq:PotentialDTerm}) 
takes many different shapes.  In order to obtain
inflation, we would like to engineer the potential to contain a sufficiently flat region within which slow roll
inflation can occur.
The challenge, nicknamed by \cite{DelicateUniverse} as the ``Delicate Universe," is that changing the parameters
of (\ref{eq:Parameters}) by small amounts appears to lead to drastic changes in the suitability of the potential
for inflation.

\begin{figure}[htp]
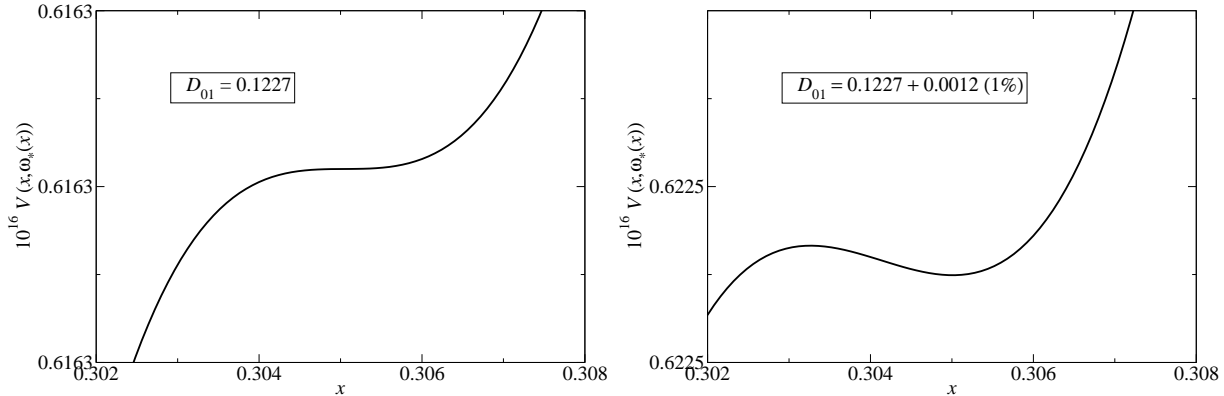

\centerline{
\includegraphics[scale=.35]{FineTuning1}
\includegraphics[scale=.35]{FineTuning2}}
\caption{\small The inflationary potential as a function of the $D3$-position $x$ for the 
parameters given in (\ref{eq:ParametersFineTune}) and $D_{01}$ modified
by $1\%$. The potential is ``delicate" since changing the 
value of $D_{01}$ by $1\%$ modifies the form of the
potential from a flat inflection point to one with a local minimum.}
\label{fig:FineTuning}
\end{figure}

For example, consider the set of parameters \cite{DBraneFineTuning},
\begin{eqnarray}
\label{eq:ParametersFineTune}
&&\{D_{01}=0.1227,\omega_F=15,A_0=1,\phi_\mu=0.2406\}  \\
&& \{n=8,N_{D3}=1\}\, .\nonumber
\end{eqnarray}
The potential for these parameters, shown in the left hand side of
Figure \ref{fig:FineTuning}, has an 
inflection point near $x\approx 0.305$ which can support inflation.  Now let us 
change one of the parameters, $D_{01}$, by $1\%
$ to 
$D_{01} = 0.1239$.  The resulting potential, shown in the right hand side of
Figure \ref{fig:FineTuning}, no longer supports
inflation due to the presence of a local minimum.
Thus, in order to obtain a model which supports inflation, for the values
of the parameters we chose we must tune $D_{01}$ to, at least, less than
the $1\%
$ (1 part in 100) level.

This simple one-dimensional fine tuning illustrates the basic problem of the ``Delicate Universe."
A more comprehensive scan of the entire parameter space has been carried out in \cite{DBraneFineTuning},
with the conclusion that the set of optimal parameters only needs to be tuned {\it by hand} to the level of one part in two.
In Section \ref{sec:Dynamical}
we will show that this model
actually contains a mechanism that can {\it dynamically} tune the parameter $D_{01}$ 
at the level of less than $10\%
$, removing the need for the arbitrary fine tuning by hand
for this parameter (the other parameters still need to be tuned by hand at the levels
discussed in \cite{DBraneFineTuning}, or tuned dynamically by some other mechanism).

\subsection{Overshoot Problem}

A generic weakness of small field inflation models, and in particular of
inflection-point potentials which arise in the model we study here,
is that the initial conditions for the inflaton's position and momentum
must be fine tuned in order for inflation to occur.  In particular,
for an inflection-point potential if the inflaton starts near the inflection
point with too large of an initial momentum (or starts at an initial condition
high up on the non-slow roll part of the potential) then it will be moving too
fast along the flat part of the potential for slow-roll inflation to occur.
In order to prevent this overshooting of the inflationary region we
must fine tune the initial conditions by hand.

There is a technical complication in studying the overshoot problem
for our model concerning the angular coordinates
of the mobile D3 brane in the warped throat.  As was shown in
\cite{DelicateUniverse}, the position of the angular minimum shifts suddenly
when the radial coordinate $x$ passes some critical value $x_c$.  This
was determined in Appendix C of \cite{DelicateUniverse} to be the value at
which the quantity $X$ in eq.\ (C.24) changes sign. We generalize the equation for $X$ to $N_{D3}>1$:
\begin{eqnarray}
X &= &\pm{2{\cal C}N_{D3}^{1/4}\over n}\frac{(x/\sqrt{N_{D3}})^{3/2}}{\left|1\mp(x/\sqrt{N_{D3}})^{3/2}\right|^{2(1-N_{D3}/n)}}\left\{2\omega+{9\over2}-\frac{(2\omega_F+3)e^{\omega-\omega_F}}{\left|1\mp(x/\sqrt{N_{D3}})^{3/2}\right|^{N_{D3}/n}}\right.\nonumber\\
&&\left.\mp\frac{3(1-N_{D3}/n)}{\left|1\mp(x/\sqrt{N_{D3}})^{3/2}\right|^2}\left[\left({x\over\sqrt{N_{D3}}}\right)^{3/2}\left(1\mp\left({x\over\sqrt{N_{D3}}}\right)^{3/2}\right)-c{x\over\sqrt{N_{D3}}}\right]\right\}.
\end{eqnarray}
For the model
parameters we are primarily interested in (see below), the critical value is close
to $x_c=1.59$.  

\begin{figure}[htp]
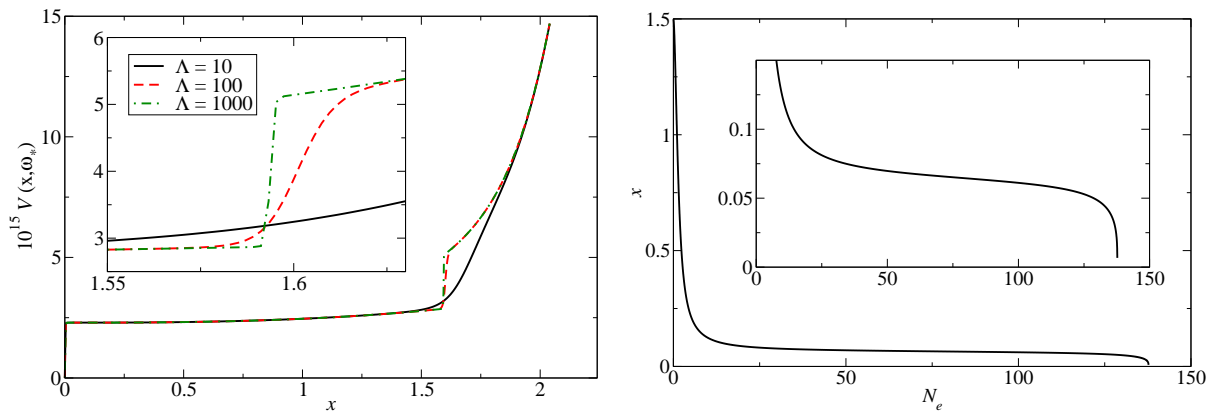

\centerline{
\includegraphics[scale=.36]{VLambda}\hskip12pt\includegraphics[scale=.35]{Ne-x}}
\caption{\small Left: potential with smoothed discontinuity, for
smoothing parameter 
$\Lambda = 10,\ 100,\ 1000$ (see eq.\ (\ref{smoothing})).  Inset shows close-up of the transition
region where angular brane moduli suddenly change.
 Right: The inflaton $x$ as a
function of number of e-foldings.  Inset is a close-up of the
small-$x$ region where most of inflation occurs.}
\label{fig:Potential}
\end{figure}

Instead of following the detailed motion of the angular brane moduli,
we will assume that they quickly move to their new minima since they are
heavy.  This leads to a sudden drop in the potential, and if this 
process happened instantaneously, it would make $V(x)$ discontinuous
at $x_c$.  The discontinuity shows up in $V$ through a sign change in
various terms in the F-term potential, namely
\begin{eqnarray}
V_F(x,\omega) &=& \frac{a|A_0|^2}{3 U^2(x,\omega)} e^{-2\omega} 
g_\pm(\frac{x}{\sqrt{N_{D3}}})^{2N_{D3}/n}\nonumber\\
&&\times\left[2\omega+6-2 (2\omega_F+3)e^{\omega-\omega_F}g_\pm(\frac{x}{\sqrt{N_{D3}}})^{-N_{D3}/n} \right. \nonumber \\
&& \left. + \frac{3N_{D3}}{n g_\pm(\frac{x}{\sqrt{N_{D3}}})}
 \left(\frac{c x}{N_{D3}^{1/2}g_\pm(\frac{x}{\sqrt{N_{D3}}})}\mp (\frac{x}{N_{D3}^{1/2}})^{3/2}\right)\right]
\end{eqnarray}
where $g_\pm(x) = 1\pm (x/\sqrt{N_{D3}})^{3/2}$.  To smooth this out, we make the 
replacement 
\be
\pm 1\to \tanh(\Lambda(x_c - x))
\label{smoothing}
\ee
where $\Lambda$ is
chosen to be sufficiently large so that the potential for $x<x_c$
has reverted to its usual form (\ref{eq:PotentialFTerm}) within
a few Hubble times after $x$ passes $x_c$.  We show the form of
the potential for several values in Figure \ref{fig:Potential}.
The magnitude of the vertical shift in the potential is relatively 
small, and it occurs 
far away from the inflationary region $x\sim 0.06$, where the
inflection point is located.  We take $\Lambda=100$,
which insures that the effects of the transition have damped out
within a few e-foldings after $x$ passes $x_c$.  

\begin{figure}[htp]
\centerline{
\includegraphics[scale=.45]{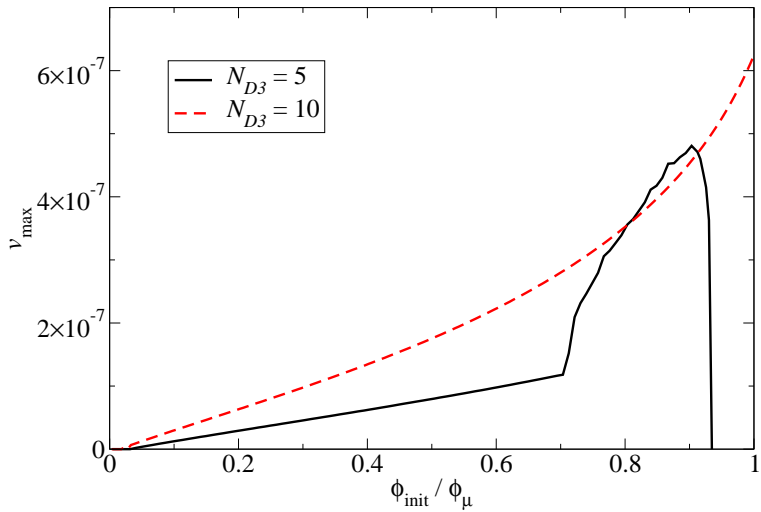}}
\caption{\small 
The allowed initial positions $\phi_{\rm init}/\phi_\mu$ and velocities $v_{\rm init}$
in the $-x$ direction which give rise to $N_e \geq 60$ are in the
region below the curves $\nu_{\rm max}$, which is defined as the maximum
allowed velocity before overshooting.  The solid curve corresponds
to the set of parameters (\ref{optimal}) which lead to a flat inflection
point.  The dashed curve corresponds to $N_{D3}=10$ which
leads to a local minimum.  The trapping of the inflaton by the local
minimum occurs for a large range of initial conditions.
}
\label{fig:Maxv}
\end{figure}

With this modification we are able to explore the effects of starting high
up on the potential, where it is much steeper and the overshoot
problem becomes evident.  For a given initial field
value $x_{\rm init}$ and momentum in the $-x$ direction $v_{\rm init}$
it is straightforward to follow the evolution of the field and
determine the total number of e-folds.\footnote{Due to the large friction, initial velocities in the positive direction will be damped away and apparently no fine tuning in this direction is needed.}  
Since the potential
diverges as $x\rightarrow \sqrt{N_{D3}}$ when the $D3$-brane approaches the
$D7$-brane, we must restrict the initial field values to
$0\leq x \leq \sqrt{N_{D3}}$. In fact
one must sometimes be more restrictive, taking
 $x\le{\tt min}(\sqrt{N_{D3}},x_{\omega_*})$, 
where $x_{\omega_*}$ is the maximum value for which
 $\partial V/\partial \omega=0$ has a solution, {\it i.e.}, 
for which there exists a stable trajectory.
To be concrete, let us
consider the optimal parameter set (the set of parameters with the 
smallest amount of overall relative fine tuning)
for which the potential has a flat inflection
point which supports inflation:\footnote{Ref.\ \cite{DBraneFineTuning} only considers $N_{D3}=1$; here we present an optimal parameter set with $N_{D3}=5$.}
\begin{equation}
  D_{01} = 0.1995,\quad \omega_F = 9.761,\quad A_0 = 0.02058,\quad \phi_\mu = 0.5804, \quad n = 8,\quad N_{D3}=5\, .
\label{optimal}
\end{equation}
The allowed initial conditions which lead to $N_e \geq 60$
e-folds of inflation are shown as the region below the solid
line in Figure \ref{fig:Maxv}.
For $\phi_{\rm init}/\phi_\mu>0.93$, the trajectory is unstable and
$\nu_{\rm max}$ is set to zero; for any given initial field value 
$\phi_{\rm init}/\phi_\mu\le 0.93$ there is always a maximum velocity
$v_{\rm init} \leq \nu_{\rm max}(x_{\rm init})$ for which inflation occurs.

Naturally, if the potential has a local minimum rather than an inflection
point then we expect that the inflaton will become trapped
in the local minimum if it is moving sufficiently slowly.  If the inflaton
starts with too much initial momentum or too high on the potential, however, then
it can still overshoot the local minimum.  The region below the dashed line of
Figure \ref{fig:Maxv} corresponds to the initial conditions for which
the inflaton is trapped in the local minimum of the potential with parameters
(\ref{optimal}) except with larger $N_{D3} = 10$.

\begin{figure}[htp]
\centerline{
\includegraphics[scale=1]{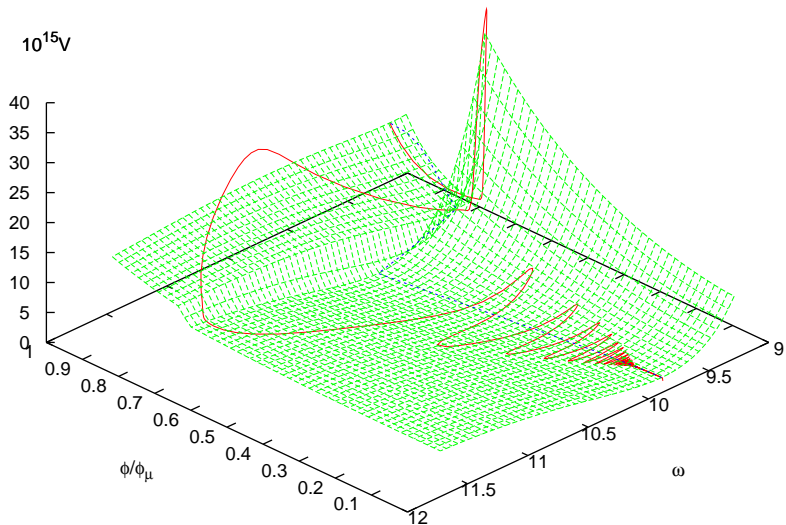}}
\caption{\small 
The trajectory (solid line) of the inflaton with initial conditions $\phi_{\rm init}/\phi_\mu=0.91$ and $v_{\rm init}=\nu_{\rm max}(\phi_{\rm init})$. The dashed line is the trajectory of the instantaneous minimum; see (\ref{eq:w*}).}
\label{fig:traj_3D}
\end{figure}

One might imagine that the inflaton will overshoot
the inflection point if it rolls down the steep part of the potential
where the transition happens at $x_c$. The solid line in Figure
\ref{fig:Maxv}, however, belies this expectation, since $x_c=1.59$
corresponds to $\phi/\phi_\mu = 0.71$ by eq.\ (\ref{phimu}), yet we
find successful examples of inflation up to $\phi/\phi_\mu>0.9$.
 Figure
\ref{fig:traj_3D} shows the trajectory of the inflaton starting above
the transition point. The inflaton runs into a potential barrier and
bounces back to follow the instantaneous minimum trajectory along
which inflation occurs. This process damps out a large fraction of
the kinetic energy and explains the curious bump in Figure
\ref{fig:Maxv}. A detailed investigation of the phase space
$(\phi,\omega,\dot\phi,\dot\omega)$ would be needed for fully
understanding the 
sensitivity to initial conditions, but this is beyond the scope
of the present work.
Nevertheless, it is clear that the unexpected
shift of the instantaneous minimum trajectory implies $v_{\phi}$ is
sensitive to the position of $\omega$, and hence inflation becomes
more intricate. The case of $N_{D3}=10$, however, is less sensitive
since the transition point does not occur in the region $\phi/\phi_\mu\le1$.

\subsection{Stabilization of the K\"ahler Modulus}

\begin{figure}[htp]
\centerline{
\includegraphics[scale=.45]{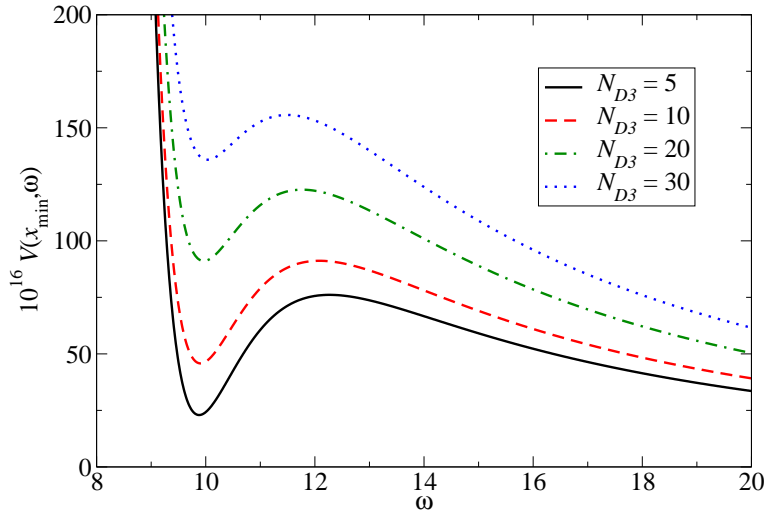}}
\caption{\small The addition of many mobile D3-branes (stabilized at the local minimum of the potential,
$x_{\rm min}$) does not destabilize the K\"ahler modulus, although it does change the stabilized value.}
\label{fig:KahlerPotential}
\end{figure}

We conclude this section with a discussion of the stabilization of the K\"ahler modulus in the scenario with
multiple $\overline{D}$-branes and uplifting sectors.
One should be concerned that with the addition of many antibranes one introduces a large amount of uplifting
energy, which could destabilize the minimum of the K\"ahler modulus.  If the uplifting energy is dominated by
other sectors, however, 
\begin{equation}
V_D(x,\omega) \approx \frac{1}{U^2(x,\omega)}\left(D_1 + 2N_{D3}T_0\right) \approx \frac{D_1}{U^2(x,\omega)}\, ,
\end{equation}
then it is clear that the stabilization of the K\"ahler modulus
is largely independent of the number of antibranes in the inflationary throat since they
only enter into the potential through this term.  In particular, in Figure
\ref{fig:KahlerPotential}, we see that the K\"ahler modulus is not destabilized by the addition of
${\mathcal O}(30)$ mobile $D3$-$\overline{D3}$ pairs, as expected.

\section{Dynamical Fine Tuning in Brane Inflation}
\label{sec:Dynamical}

In the previous section we identified two fine tuning problems with warped $D$-brane inflation models,
the ``Delicate Universe" problem of fine tuning in the potential, and the ``overshoot problem" of fine
tuning in the initial conditions.
Both of these problems are present in usual small field inflation as well, but are enhanced
due to the inflection-point type potential which occurs in these brane inflation models
(as well as in other string-inspired inflationary models 
\cite{Itzhaki:2007nk,Itzhaki:2008ih,AccidentalInflation,Conlon:2008cj,AxionicD3D7}).
In what follows, we will describe a dynamical mechanism in D-brane inflation models
to achieve these required fine tunings.

Let us use a modified set of parameters for our model (exchanging $D_{01}$ for $T_0$),
\begin{eqnarray}
\label{eq:Parameters2}
\{T_0,\omega_F,A_0,\phi_\mu\} & & \mbox{effectively continuous} \\
\{n,N_{D3}\} & &\mbox{discrete}\, .\nonumber
\end{eqnarray}
Notice that this implies $D_{01}$ is no longer a free parameter, but rather that it depends on the
number of $\overline{D}$-branes in the throat,
\begin{equation}
D_{01} = 2\frac{N_{D3}T_0}{D_1} \equiv N_{D3} \Delta D_{01}(\omega_F,A_0,n,T_0)\, ,
\label{eq:D01def2}
\end{equation}
where we emphasized the fact that $\Delta D_{01}$ is a function of the
other parameters.
The number of $\overline{D}$-branes in the throat is not constant since
$D$-branes can annihilate with $\overline{D}$-branes at the tip of the throat.
Keeping all other parameters fixed, removing $D$-branes effectively 
decreases $D_{01}$ in steps per brane of
\begin{equation}
\Delta D_{01} = 2\frac{T_0}{D_1}\, .
\label{eq:D01StepSize}
\end{equation}

We can then imagine the following scenario: in the pre-inflationary scenario, there are a
number of $\overline{D}$- and $D$-branes in the compact space.  The $\overline{D}$-branes
are quickly attracted to the tip of warped throats by the flux-induced $D3$-charge.  The remaining
$D$-branes migrate towards the $\overline{D}$-branes, attracted by Coulomb attraction and 
moduli stabilization effects.  All of the parameters (\ref{eq:Parameters2}) are fixed
except the number of $D3$-branes $N_{D3}$ which may be large.  As a result, the parameter
$D_{01}$ in (\ref{eq:D01def2}) is large, which generically leads to a local minimum
in the potential, as shown in the right hand plot of Figure \ref{fig:FineTuning}.  As the $D$-branes
fall into the throat (with random initial conditions) they are trapped at the local minimum by two
effects: lack of enough energy to overcome the potential barrier, and loss of energy due to open
string production by collision with other $D$-branes trapped at the minimum.  Gradually, all of the mobile $D$-branes
are sitting in a false vacuum of the potential.  One by one, the $D$-branes
will tunnel out of the false vacuum and annihilate against one of the $\overline{D}$-branes,
decreasing the number of $D$-branes in the throat and subsequently decreasing the parameter
$D_{01}$ in steps of $\Delta D_{01}$.  The tunneling process continues until a sufficiently large number of $D$-branes
tunnel out of the local minimum so that $D_{01}^{(n)} = D_{01} - n \Delta D_{01}$ is equal to a critical
value at which the local minimum disappears and a flat inflection point appears.  The remaining $D$-branes
then inflate along this dynamically tuned inflection point.  
Because all of the remaining $D$-branes are subject to the same force, the stack
of branes moves uniformly with the inflaton identified as the radial position of the 
stack.
The schematic picture of this scenario
is shown in Figure \ref{fig:BranesInThroat}.

In order
to remove the necessity to fine-tune by hand the value of $D_{01}$, we would like the step size $\Delta D_{01}$
to be sufficiently small.  More precisely, let us consider a fixed
set of parameters in (\ref{eq:Parameters2}) (except for $N_{D3}$ of course),
let $D_{01}^*$ be the value of $D_{01}$ at which
a flat inflection point which supports $60$ or more e-folds occurs, and let us say that $D_{01}^*$
must be tuned to the level of $1$ part in $1/\delta$ (e.g.\ to the 
$\delta\times 100\%$ level).  
In order to dynamically fine tune the scenario, we need
the steps as $D$-branes tunnel from the minimum to be at least equal to the amount of tuning required,
e.g.\ $\delta \geq \Delta D_{01}/D_{01}^*$.  
{}From (\ref{eq:D01def2}) this implies that we need at least
\begin{equation}
N_{D3} \geq \delta^{-1}
\end{equation}
$D$-branes initially in order to achieve the necessary fine tuning.
Clearly, models which require a significant amount of tuning also require a significant number
of initial mobile $D$-branes, at which point the probe approximation for the mobile $D$-branes
may break down.  The point at which the probe approximation breaks down depends on the effective
$D3$-charge of the fluxes which generate the warped throat (more precisely,
the condition depends on the backreaction of the mobile $D$-branes on the curvature
scale of the warped throat), $N_{D3} < N_{\rm eff,D3}$.  For typical values up to the bound
$N_{\rm eff,D3}\sim {\mathcal O}(10^6)$, we see that we cannot use this mechanism to address severe
fine tuning of $D_{01}$ at the level of $1$ part in $10^6$ or more.  In a later subsection we will actually
find that the requirement that the dynamical flattening tunneling process be faster than
the timescale for annihilation of the anti-branes with the flux in the throat \cite{KPV}
forces the number of antibranes to be small; $N_{D3} \sim {\mathcal O}(10)$, corresponding to
a dynamical fine tuning at the level of $1$ part in $10$ or so, is a more reasonable expectation
(although it may be possible to relax this slightly by allowing different numbers of branes
and anti-branes).

\subsection{Concrete example}
\label{concrete}

We will now describe a specific set of parameters in which the dynamical mechanism described above
does lead to a dynamical scanning of the parameter space ending in a viable inflationary
model which matches observational constraints.
Suppose that inflation takes place for the optimal parameter values (\ref{optimal}), which 
corresponds to taking the following values for the microscopic
parameters:
\begin{equation}
 Q_{\mu} = 1.07,\ N_5=10,\ B_4 = 76.95,\ B_6 = 1.0370
\end{equation}
and the corresponding value of $D_1$ is $7.260\times
10^{-12}$.
This set of parameters requires  the same degree of fine tuning 
as the optimal set shown in ref.\ \cite{DBraneFineTuning}: $D_{01}$
must be tuned at the 20\%
level ($\delta D_{01}\cong 0.04$) in order to get successful inflation.
To ensure passing  through the experimentally allowed interval,
the uplifting sources must be chosen such that the increment in $D_{01}$ as branes tunnel from the
stack is $\Delta D_{01}\cong 0.04$.  Thus
with the choice of antibrane tension $T_0 = 1\times 10^{-13}$,
which by the relation $T_0 \sim h_i^4 m_s^4/g_s$ implies that the warp
factor at the tip of the throat is approximately $h_i \sim 10^{-2}$
(assuming $m_s \approx 0.01 M_p$),
we have the step size
\begin{equation}
	\Delta D_{01} = 2 {T_0\over D_1} = 0.04.
\label{eq:StepSizeEx}
\end{equation}
Since $D_{01} = N_{D3} \Delta D_{01}$, there must be $N_{D3}=5$
branes in the stack at the time inflation takes place.  

\begin{figure}[htp]
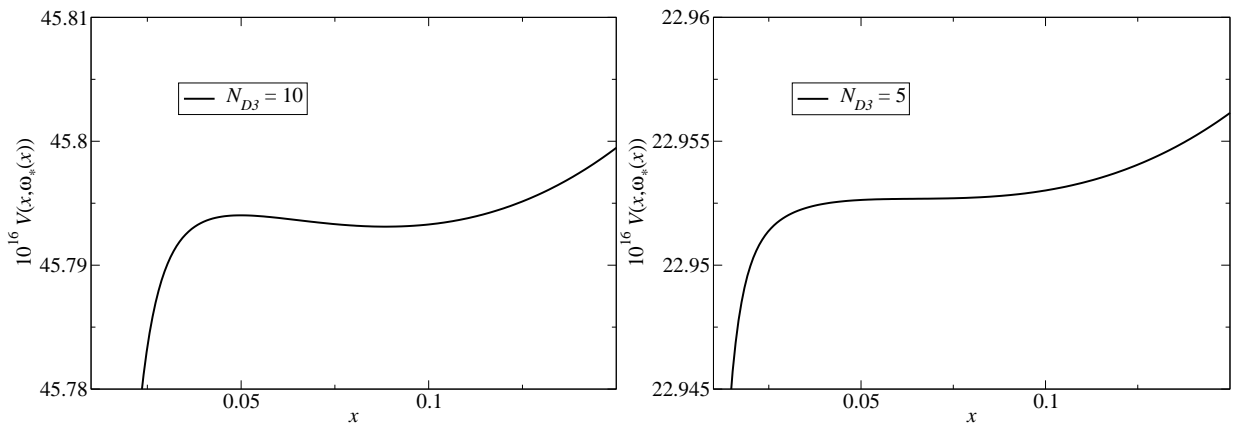

\centerline{\includegraphics[scale=.37]{V10_N5}\includegraphics[scale=.37]{V5_N5}}
\caption{\small For a large number of mobile D3-branes a local 
minimum in the inflationary potential exists,
trapping the branes at a finite distance in the throat.  As individual branes tunnel out of the local minimum,
the potential barrier decreases until the local minimum vanishes and is replaced by a very flat inflection point
potential.  (The height of the potential also goes down with the
number of branes.)
The tunneling of the branes leads to the dynamical tuning of 
the parameter $D_{01}$ in small steps.}
\label{fig:DynamicalFlattening}
\end{figure}

For concreteness, imagine starting with $N_{D3} = 10$ mobile
$D3$-branes, with a corresponding value of $D_{01} = 0.3990$. The
potential, shown in Figure \ref{fig:DynamicalFlattening}, clearly has
a local minimum around $x\approx 0.09$.  
As the number
of $D3$-branes is decreased (corresponding to decreasing $D_{01}$ in steps of
$\Delta D_{01}$ (\ref{eq:StepSizeEx})),  we find that the local
minimum disappears and is replaced by an inflection point when the
number of mobile $D3$-branes is $N_{D3} = 5$, corresponding to the 
value $D_{01} = 0.1995$.  Subsequently following the evolution of the
two-scalar field system $(x,\omega)$ as in \cite{DBraneFineTuning},
we find an inflationary power spectrum with

\begin{eqnarray}
{\mathcal P}_R &=& 2.41\times 10^{-9} \\
n_s &=& 0.989
\end{eqnarray}
which is within the $2\sigma$ confidence level of the latest WMAP observations.

\subsection{Discussion}

We have seen an example where the range of values for $D_{01}$
leading to a successful amount of inflation is expanded by our dynamical 
tunneling
mechanism from its naive extent of ``by hand" tuning, $\delta  D_{01} \cong 0.04$, to a
much larger range  $\delta  D_{01} \cong 0.2$.  This is in fact
as large as the actual value $D_{01} \cong 0.2$ needed during the
final stage after all tunneling events have finished and slow-roll
inflation begins; thus the residual fine-tuning problem for this
parameter is completely eliminated.  It is especially
interesting that the mechanism dynamically tunes $D_{01}$, since
this is the parameter to which the inflaton potential was found to
be most sensitive in  \cite{DBraneFineTuning}, for the purpose of
getting successful inflation.

\begin{figure}[htp]
\centerline{\includegraphics[scale=.5]{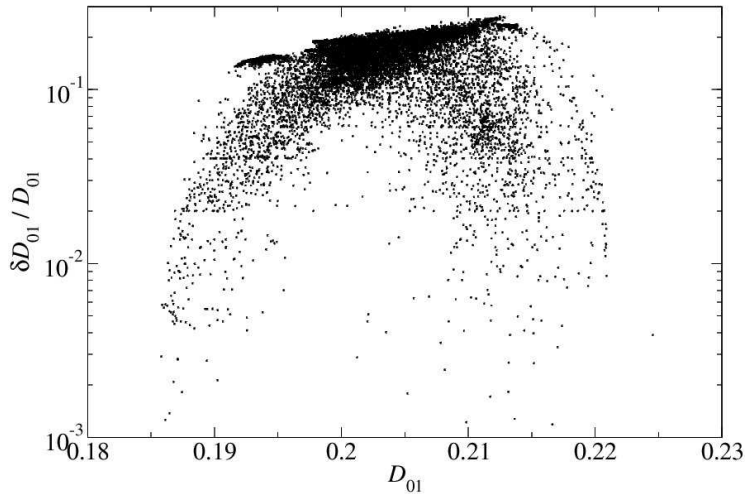}}
\caption{\small Scatter plot of 
the fractional amount of  required  fine tuning $\delta D_{01}/D_{01}$
versus $D_{01}$
for approximately $22,000$ sample inflationary solutions.  Most models
only require tuning at the level of $1$ part in $10$.}
\label{fig:deltaD01}
\end{figure}

Generalizing the above example, we see that in order for this dynamical
mechanism to work the change in $D_{01}$
due to tunneling, $\Delta D_{01}$, must be no greater than the naive
``by hand" tuning
required for successful inflation, $\delta D_{01}$. 
If this is true, we are
guaranteed to pass through the allowed range during some tunneling
transition, provided that $D_{01}$ is initially greater than the
desired value for inflation.  In Figure \ref{fig:deltaD01} we show
the results of a scan over a large number of models, determining the value of the
parameter $D_{01}$ which gives rise to successful inflation
and the fractional amount of fine tuning $\delta D_{01}/D_{01}$ of the parameter
which is required.  It is clear that most of the models require
no more tuning than $1$ part in $10$, which can easily be done
dynamically with the mechanism presented here, although it is 
possible that certain individual models may require more tuning than
this.

The effective range of allowed values
for $D_{01}$ becomes $\Delta N_{D3}\Delta D_{01}$, where $\Delta N_{D3}$
is the number of
branes which tunnel between the initial configuration and the final
one at which the local minimum disappears and slow-roll inflation
begins.
This effective range is greater than the naive range of ``by hand" tuning
by the factor $\Delta N_{D3} \Delta D_{01}/\delta D_{01}$.  
In the  example above, we took $\Delta D_{01}= \delta
D_{01}$ and  $\Delta N_{D3}=5$ to find an enhancement by a factor of
$5$, but it is possible to do even better by taking a larger initial
number of branes.  This number is only limited by the
destabilization of the K\"ahler modulus and the
danger of back-reaction of the branes distorting the background
geometry in a way which is not taken into account.  For the latter, if the
back-reaction effects preserve the existence of a local minimum in the potential
the dynamical tunneling regime will still
occur, and our mechanism is rather insensitive to the details of the
tunneling events as long as normal inflation occurs in the end.  

Even if the model parameters do not obey the condition
$\Delta D_{01}\le \delta D_{01}$, we still obtain an enhancement in
the range of parameters for which successful inflation can occur, only
in this case it is a discrete series of islands in the space of
$D_{01}$ rather than a nearly continuous range.  As long as $D_{01}$ exceeds
the value needed for inflation and $D_{01}/\Delta D_{01}$ is within
$\delta D_{01} / \Delta D_{01}$ of being an integer, the last
tunneling event will lead to slow roll inflation with the desired
properties.  

In addition to widening the parameter space compatible with inflation,
the space of viable initial conditions for multiple branes is augmented.  
One effect is that the presence of the local minimum
for large numbers of $D3$-branes makes it more difficult to overshoot.
Simply comparing the areas below the curves in Figure \ref{fig:Maxv}
we see that the local minimum increases the allowed
volume of initial conditions leading to $60$ e-folds by a factor of 2.
This is actually an underestimate of the effect, since loss of energy
through collisions between mobile $D$-branes and other $D$-branes already
trapped at the local minimum can be a significant effect 
\cite{BeautyAttractive,McAllister:2004gd}.
We also saw in Figure \ref{fig:traj_3D}
that in the presence of additional branes
the multifield trajectory becomes more intricate, and this can
lead to additional helpful features (such as the barrier we saw there)
that reduce the overshooting problem.  It is clear from this that a
more systematic study of the multifield initial conditions space
is needed to say more about the tendency of these systems to overshooting,
but it also seems that the tendency for overshooting is decreased when
the effects of multiple branes are included.

\subsection{Tunneling Rate}

In the scenario as we have so far described it, the inflationary
brane stack can sit in the metastable local minimum for an arbitrary
amount of time before the final tunneling which initiates
normal slow-roll inflation.
However, there is a competing tunneling process which could in
principle interfere with this picture, namely  the annihilation of
the $\overline{D}$-branes at the tip of the throat with the
background flux, as described in ref.\ \cite{KPV} (KPV).  We need to make sure
that the branes in our stack have time to tunnel before this
geometry-changing transition can take place.  To this end, we
estimate the rates for the two processes in the present section.  

First, let us estimate the rate for $D$-branes to tunnel from the
local minimum.  We note that the potential does not flatten out
immediately after a $D$-brane tunnels, but rather it only flattens
once the $D$-brane classically rolls to the tip of the throat and
annihilates with an $\overline{D}$-brane.  This means that there is
no energy benefit for multiple branes to tunnel at once, so that the
rate for $n$ branes to tunnel is the $n$th power of the
tunneling rate for one brane; thus single brane tunneling is
most likely.  Further, because the tunneling timescale is long, any
velocity-dependent interactions between the branes, such as
open-string creation, are suppressed as well, and we
can use a single field description for tunneling.

There are several different estimates of the tunneling rate, depending
on the features of the potential.  The rate for standard 
Coleman-De Luccia (CDL) instantons depends
on whether there is significant gravitational backreaction \cite{CDL},
\begin{equation}
\Gamma_{CDL} \sim \cases{
	\mbox{exp}\left(-\frac{27\pi^2}{2} \frac{T^4}{V_{\rm min}^3}
\right), & w/o\, gravitational back-reaction ($M_p^2 V_{\rm min} \gg T^2$) \cr
	\mbox{exp}\left(-\frac{24\pi^2}{V_{\rm min}}\right), & 
with gravitational back-reaction ($M_p^2 V_{\rm min} \ll T^2$)\,, \cr}
\end{equation}
where $V_{\rm min}$ is the vacuum energy in the false vacuum, and the
tension of the bubble of true vacuum
\begin{equation}
T = \int d\phi \sqrt{2 V(\phi)} \sim \sqrt{V_{\rm max}} \Delta \phi
\end{equation}
depends on the height $V_{\rm max}$ and width $\Delta \phi$ of the potential 
barrier.
Alternatively, Hawking-Moss instantons depend only upon the separation
between the false vacuum and the maximum of the potential 
barrier\footnote{The original estimate in \cite{HM} appears
to have missed a factor of 3 in the bounce action.} \cite{HM},
\begin{equation}
\Gamma_{HM} \sim \mbox{exp}\left(-24\pi^2\left(\frac{M_p^4}
{V_{\rm min}}-\frac{M_p^4}{V_{\rm max}}\right)\right)\, .
\end{equation}

In the dynamical fine tuning scenario outlined above, the height of the
potential barrier is small and decreases with each successive tunneling
event, so we have $V_{\rm max} = V_{\rm min}(1+\epsilon)$ for some $\epsilon \ll 1$.
We can ignore gravitational backreaction in the CDL instanton estimate since
\begin{equation}
\frac{T^2}{M_p^2 V_{\rm min}} \sim \frac{\Delta \phi^2}{M_p^2}\frac{V_{\rm max}}{V_{\rm min}} 
	\sim \frac{\Delta \phi^2}{M_p^2} \ll 1\,,
\end{equation}
where in the last step we used that $\Delta \phi/M_p \ll 1$ for warped
brane inflation models \cite{Chen:2006hs,BaumannMcAllister}.
The decay rate for branes to tunnel from the local minimum through
the dynamical fine tuning process is then
\begin{equation}
\Gamma_{\rm dyn} \sim \mbox{exp}\left(-24\pi^2 \frac{\alpha M_p^4}{V_{\rm min}}\right),
\label{Gdyn}
\end{equation}
where $\alpha$ is the smaller of 
$\{\frac{9}{16} \frac{\Delta \phi^4}{M_p^4},\epsilon\} \ll 1$. 
For typical shapes of the potential we have $\Delta \phi^4/M_p^4 \ll \epsilon$,
so the decay rate is dominated by the CDL instanton.

Next we turn to the 
other relevant quantum tunneling process\footnote{In principle one should
also consider the decay of the dS minimum to the runaway Minkowski 
vacuum in the K\"ahler modulus direction, but this decay is dominated by
CDL instantons {\it with} gravitational backreaction and is much more
suppressed than the dynamical tunneling or KPV tunneling rates.} for this scenario,
that of the annihilation of the $\overline{D3}$-branes with the flux
in the throat, as described by Kachru, Pearson and Verlinde (KPV) \cite{KPV} and
studied in greater detail in the references
\cite{Frey:2003dm,Freivogel:2008wm,Brown:2009yb}.  With $M$
units of $F_3$ flux wrapped on the $A$-cycle of the throat
and $K$ units of $H_3$ flux wrapped on the $B$-cycle of the throat
we have an effective $D3$-charge of $N_{\rm eff} = MK$ and a warp factor
of $h_A \sim e^{-2\pi K/(3g_s M)}$.  When there are $p$ $\overline{D3}$-branes
at the tip of the throat with $p/M \geq 0.08$, the configuration is 
classically unstable against the $\overline{D}$-branes annihilating with
the fluxes.  Since we have the same number of branes and antibranes,
this puts another limit on the number 
of $D$-branes we can have in our compactification, $N_{D3} \leq
0.08\times N_{\rm eff,D3}/K$
which is somewhat more stringent than the one derived before due to backreaction.
We will assume that we have large enough flux $M$ so that this effect can
only proceed via quantum decay through CDL instantons.  The rate,
including corrections found in \cite{Freivogel:2008wm}, is
\be
	\Gamma_{KPV} \sim \mbox{exp}\left(-\beta\,{g_s M^6\over N_{\overline{D3}}^3}\right)\,,
\ee
where $\beta = 3\times 10^{-3}$ is a pure number fixed for the geometry.
Demanding that this be slower than the brane-tunneling rate
(\ref{Gdyn}) gives us a restriction on the $F_3$ flux,
\be
	M > \left(\frac{27\pi^2}{2 \beta}\right)^{1/6} \frac{\Delta x^{2/3} \phi_\mu^{2/3} N_{D3}^{1/2}}{g_s^{1/6} V_{\rm min}^{1/6}}
\ee
(where we have again used our simplifying assumption that 
$N_{\overline{D3}} = N_{{D3}}$).
Using the definition $\phi_\mu/M_p \equiv \frac{2}{\sqrt{MK}} \frac{1}{Q_\mu \sqrt{B_6}} < \frac{2}{\sqrt{MK}}$ where $Q_\mu,B_6 \geq 1$
and the definition of the vacuum energy in the local minimum $V_{\rm min} = \frac{2 N_{D3} h_A^4}{(2\pi)^3 g_s \ell_s^4}$ 
in terms of the fluxes $h_A = e^{-2\pi K/(3 g_s M)}$, we can turn the requirement
that KPV tunneling be suppressed into a bound on the parameter $\phi_{\mu}$,
\begin{equation}
\phi_\mu < \phi_{\rm max}\equiv 2^{3/5} \left(\frac{8
\pi}{3}\right)^{3/10} \left(\frac{2\beta}{27\pi^2}\right)^{1/10}
\frac{(V_{\rm min}/M_p^4)^{1/10}}{g_s^{1/5}\, \Delta x^{2/5}\,
N_{D3}^{3/10} \left(\ln\frac{2N_{D3}}{(2\pi)^3 V_{\rm min} g_s \ell_s^4}\right)^{3/10}} \, .
\label{phiUP}
\end{equation}
When this bound is satisfied, KPV tunneling will be suppressed relative to 
the dynamical tuning effect.
Surprisingly, this bound is difficult to satisfy---for the ``optimal" parameter
set presented in (\ref{optimal}) with
$$
V_{\rm min} \cong (2.6\times 10^{-4})^4,\, \Delta x \cong 0.05,\, N_{D3} \cong 10,\, g_s \cong 0.1,\, \ell_s \cong 0.1,
$$ 
we have $\phi_{\rm max} \approx 0.025$, which is much smaller than the value of the parameter $\phi_{\mu} = 0.5804$.  
One way to satisfy the bound is to adjust $\phi_{\rm max}$ by reducing the value of 
$g_s$; however, since the bound (\ref{phiUP}) is only
sensitive to $g_s^{1/5}$ we need to reduce the string coupling by a large
factor to $g_s \sim 10^{-8}$.  This allows the bound to
be satisfied $\phi_{\rm max} \approx 0.70 > \phi_{\mu}$ and leads to the tunneling rates
\begin{eqnarray}
\Gamma_{\rm dyn} &\sim & e^{-2\times 10^{10}} \nonumber \\
\Gamma_{KPV} &\sim & e^{-9\times 10^{10}} \nonumber
\end{eqnarray}
which indicate that the dynamical fine tuning effect indeed dominates.  

It is possible that there exists a parameter regime where the dominance of dynamical
flattening over KPV tunneling can be achieved with a more realistic value
of the string coupling (e.g.\ $g_s\sim 0.01$).  In particular, instead
of attempting to make the value of $\phi_{\rm max}$ in the bound larger by
suppressing $g_s$, one can search for inflationary models with small $\phi_{\mu}$.
Based on the observation that there is a degeneracy between $\phi_{\mu}$ and the
number of D7-branes $n$ \cite{DBraneFineTuning}, we have done some simple
numerical searches of the parameter space and find that most models seem
to require an equally unrealistic number of D7-branes $n\sim 10^3$, although
we have not done a complete scan of parameter space.  Since models which allow
dynamical fine tuning to dominate the decay rate do not seem to be generic 
in parameter space, this raises
the open question as to how much the dynamical effect will help the overall fine tuning
required to obtain a viable model.

\section{Conclusion}
\label{sec:Conclusion}

The idea that our universe could have emerged from a series of
tunneling events has become rather popular in the context of the
string theory landscape of vacua.  In this paper, we have provided
an explicit open-string example of this idea within the warped
brane-antibrane inflation model, where branes from a stack trapped in
the throat sequentially tunnel through a potential barrier to the
bottom of the throat.  The remarkable feature here is that it can be
natural for the final tunneling event to lead to slow roll inflation,
because the barrier becomes increasingly shallow after each
tunneling.  The concept of naturalness is quantified by the range of
values of the parameter $D_{01} = D_0/D_1$ which (together with
appropriate values of other parameters) are compatible with the CMB
power spectrum observed in our universe.  (The parameter $D_{01}$ is
the ratio of uplifting due to antibranes in the inflationary throat
versus that coming from other sources, such as other throats.)  The
range which leads to successful inflation after tunneling depends on
how many branes tunnel, and could be enhanced relative to the usual
value by a factor of 100 or even 1000, limited only by the number of
branes which can be trapped in the throat before their back-reaction
seriously alters the background throat geometry, or destabilization
of the K\"ahler modulus from its dS vacuum.

Not only can successful inflation result from a much larger range of
parameters of the model than previously thought, but also the range
of initial conditions is expanded.  This is because the problem of
the inflaton overshooting the inflection point, where inflation
should take place, is ameliorated if the inflection point is
initially replaced by a local minimum of the potential.  Once the
stack of branes is trapped at the minimum, it will naturally start
rolling away  from the flattest region of the potential with
initially vanishing velocity.  This occurs  at the moment of the
final tunneling event when the last shallow local minimum converts to
a monotonic potential, which is close to having an inflection
point.

In order for the dynamical tunneling process to be faster than
other decay processes, such as $\overline{D3}$-branes annihilating
with the flux at the tip of the throat as in KPV \cite{KPV}, we found
we needed to tune the string coupling to be quite small, 
although other parameter regimes without such extreme tuning of
the string coupling may exist.  It remains
to be seen how generic this picture of a dynamical open string
inflationary landscape is within the full set of allowed parameters.

\section*{Acknowledgments} We would like to thank A. Frey, D. Green, 
and A. Maloney for helpful discussions.
B.U.\ is supported in part through an IPP 
(Institute of Particle Physics, Canada) Postdoctoral Fellowship, and
by a Lorne Trottier Fellowship at McGill University.  L.H.\ is
supported by Carl Reinhardt Fellowship at McGill University.  Our
work is also supported by NSERC (Canada).

\bibliography{dynamicTuning}

\end{document}